\documentclass{article}

\usepackage{arxiv}

\usepackage[utf8]{inputenc} 
\usepackage[T1]{fontenc}    
\usepackage{hyperref}       
\usepackage{url}            
\usepackage{booktabs}       
\usepackage{amsmath}
\usepackage{amsfonts}       
\usepackage{nicefrac}       
\usepackage{microtype}      
\usepackage{lipsum}
\usepackage{graphicx}
\usepackage{listings}
\usepackage{xcolor}
\usepackage{algorithm}
\usepackage{algorithmic}
\usepackage{subcaption}
\usepackage{babel}

\definecolor{codegreen}{rgb}{0,0.6,0}
\definecolor{codegray}{rgb}{0.5,0.5,0.5}
\definecolor{codepurple}{rgb}{0.58,0,0.82}
\definecolor{backcolour}{rgb}{0.95,0.95,0.92}

\lstdefinestyle{mystyle}{
  backgroundcolor=\color{backcolour}, commentstyle=\color{codegreen},
  keywordstyle=\color{magenta},
  numberstyle=\tiny\color{codegray},
  stringstyle=\color{codepurple},
  basicstyle=\ttfamily\footnotesize,
  breakatwhitespace=false,         
  breaklines=true,                 
  captionpos=b,                    
  keepspaces=true,                 
  numbers=left,                    
  numbersep=5pt,                  
  showspaces=false,                
  showstringspaces=false,
  showtabs=false,                  
  tabsize=2
}
\graphicspath{ {./images/} }

\title{Monte Carlo Analysis of Boid Simulations with Obstacles: A Physics-Based Perspective}

\author{
 Quoc Chuong Nguyen \\
  Department of Mathematics\\
  University at Buffalo\\
  New York, NY 14260 \\
  \texttt{quocchuo@buffalo.edu}
}

\begin{document}
\maketitle
\begin{abstract}
Boids, developed by Craig W. Reynolds in 1986, is one of the earliest emergent models where the global pattern emerges from the interaction between many individuals within the local scale. In the original model, Boids follow three rules: separation, alignment, and cohesion; which allow them to move around and create a flock without intention in the empty environment. In the real world, however, the Boids' movement also faces obstacles preventing the flock's direction. In this project, I propose two new simple rules of the Boids model to represent the more realistic movement in nature and analyze the model from the physics perspective using the Monte Carlo method. From those results, the physics metrics related to the forming of the flocking phenomenon show that it is reasonable to explain why birds or fishes prefer to move in a flock, rather than sole movement. Code for reproducing my experiments can be found at \href{https://github.com/ChuongQuoc1413017/Multi\_Agent\_Simulation}{https://github.com/ChuongQuoc1413017/Multi\_Agent\_Simulation}.
\end{abstract}


\section{Introduction}
\label{sec:intro}

First developed by Craig W. Reynolds in 1986 \cite{10.1145/37402.37406}, the Boids model has become popular since then in the study of emergence. In this model, the objects named Boids move independently and finally form large collective flocks without intention. The name "Boid" corresponds to "bird-oid object", which means the bird-like object \cite{Banks2007}. Over time, many studies have been constructed to explore this model, see \cite{Yong2014, doi:10.1098/rsta.2016.0351, CLARK2012232, 10487362, Maruyama2019, https://doi.org/10.1002/cav.123}; such as individual Boid behavior research in \cite{10487362}, comparison of clustering methods for flock formulation in large-scale Boids model in \cite{Maruyama2019}, and the modified model defines the chance that a boid becomes a leader and tries to escape in \cite{https://doi.org/10.1002/cav.123}. From these studies, I proposed a new analysis to investigate further this model by introducing new rules and physics measures, specifically the flock's size and the motion of Boids. My work can be summarized as follows: 

\begin{itemize}
    \item \textbf{Proposed a new version}: Developed new rules for the Boids to avoid obstacles and move randomly.
    \item \textbf{Proposed physics metrics}: Developed metrics to analyze the Boids model in terms of physics.
    \item \textbf{Validated the natural phenomenon through Extensive Experiments}: Conducted comprehensive computational experiments for the variations of separation, alignment, and cohesion weights.
\end{itemize}

The report is structured as follows. Section \ref{sec:intro} gives a brief overview of the relevant studies and a summary of my work. Section \ref{sec:model} describes the details of a new model. Section \ref{sec:exp} presents my computational results, and some concluding remarks are made in Section \ref{sec:discuss}.

\section{Model}
\label{sec:model}

\subsection{Description}
\label{sub:description}

As in \cite{10.1145/37402.37406}, the Boids move in the vast environment following three rules as follows:

\begin{itemize}
    \item \textbf{Separation}: Each Boid tries to keep a safe distance from others to avoid overcrowding.
    \item \textbf{Alignment}: Each Boid tries to change its position corresponding to the other nearby Boid's average velocity.
    \item \textbf{Cohesion}: Each Boid tries to move towards the average position (center of mass) of nearby Boids.
\end{itemize}

To modify the original work, I included two simple rules to the model:

\begin{itemize}
    \item \textbf{Obstacles Avoidance}: Boids steer away from the obstacles (randomly placed in the environment) in their direction.
    \item \textbf{Wander Movement}: Boids have a small random movement to represent the actual different identities in nature.
\end{itemize}

The five rules are integrated into a weighted system, where each force contributes to the Boid's final velocity vector and position vector based on Newton's laws of motion. The weights assigned to each force determine the priority of each behavior. Detailed descriptions of the rules are provided in Section \ref{sub:math}.

\subsection{Mathematical formulations}
\label{sub:math}

According to Newton's laws of motion, see \cite{Sajwan2014Flocking} for the original version, the Boids' movement is updated using the formulas:

\begin{align}
    \mathbf{p}_{\text{new}} & = \mathbf{p}_{\text{old}} + \Delta t \cdot \mathbf{v}_{\text{new}} \\
    \mathbf{v}_{\text{new}} & = \mathbf{v}_{\text{old}} + \Delta t \cdot \mathbf{a} \\
    \mathbf{a} _{\text{new}} & = w_s \cdot \mathbf{a}_{\text{sep}} + w_a \cdot \mathbf{a}_{\text{align}} + w_c \cdot \mathbf{a}_{\text{coh}} \nonumber \\
    & + w_w \cdot \mathbf{a}_{\text{wan}} + w_o \cdot  \mathbf{a}_{\text{obs}}
\end{align}

For simplicity, I set \( m = w_w = \Delta t = 1 \), where \( m \) is the mass of Boids. This leads to the acting force \( \mathbf{F} = \mathbf{a} \), then the force for the motion is calculated via the accelerations as:

\begin{align}
    \mathbf{a}_{\text{sep}} & = \text{Scale}\left(\sum_{j \in R(r_{s}, i)} \frac{\mathbf{p}_i - \mathbf{p}_j}{ \max ( \|\mathbf{p}_i - \mathbf{p}_j\|, 1 )}, \, F_{\text{max}}\right) \\
    \mathbf{a}_{\text{align}} & = \text{Scale}\left[ \frac{1}{\left| R(r_{a}, i) \right|} \left( \sum_{j \in R(r_{a}, i)} \mathbf{v}_j \right) - \mathbf{v}_i , \, F_{\text{max}} \right] \\
    \mathbf{a}_{\text{coh}} & = \text{Scale}\left[ \frac{1}{\left| R(r_{c}, i) \right|} \left( \sum_{j \in R(r_{c}, i)} \mathbf{p}_j \right) - \mathbf{p}_i , \, F_{\text{max}} \right] \\
    \mathbf{a}_{\text{obs}} & = \text{Scale}\left(\sum_{k \in \text{obstacles}} \frac{\mathbf{p}_i - \mathbf{p}_k}{\|\mathbf{p}_i - \mathbf{p}_k\|}, \, F_{\text{max}}\right) \\
    \mathbf{a}_{\text{wan}} & = \mathcal{U}_{\left[-r_{w}, r_{w} \right]}: \text{uniform distribution}
\end{align}

, where

    \begin{itemize}
        \item \( R(r, i) = \{ j | \| \mathbf{p}_i - \mathbf{p}_j \| \leq r \} \): neighbors of Boid $ i $.
        \item \( r_{s}, r_{a}, r_{c}, r_{w} \) are separation, alignment, cohesion, and wander radii respectively.
        \item \( \mathbf{p}_i \): position of the current Boid.
        \item \( \mathbf{p}_j \): position of a neighboring Boid.
        \item \( \mathbf{p}_k \): Position of an obstacle
        \item \( \mathbf{v}_i \): velocity of the current Boid.
        \item \( \mathbf{v}_j \): velocity of a neighboring Boid.
        \item \( F_{\text{max}} \): maximum allowable force.
    \end{itemize}

For visualization purposes, we will scale the accelerations as:

\begin{align}
    \text{Scale}(\mathbf{f}, F_{\text{max}}) =
\begin{cases}
\mathbf{f} & \|\mathbf{f}\| \leq F_{\text{max}}, \\
F_{\text{max}} \cdot \frac{\mathbf{f}}{\|\mathbf{f}\|} & \|\mathbf{f}\| > F_{\text{max}}.
\end{cases}
\end{align}

\begin{figure}[htb!]
    \centering
    \includegraphics[width=\textwidth]{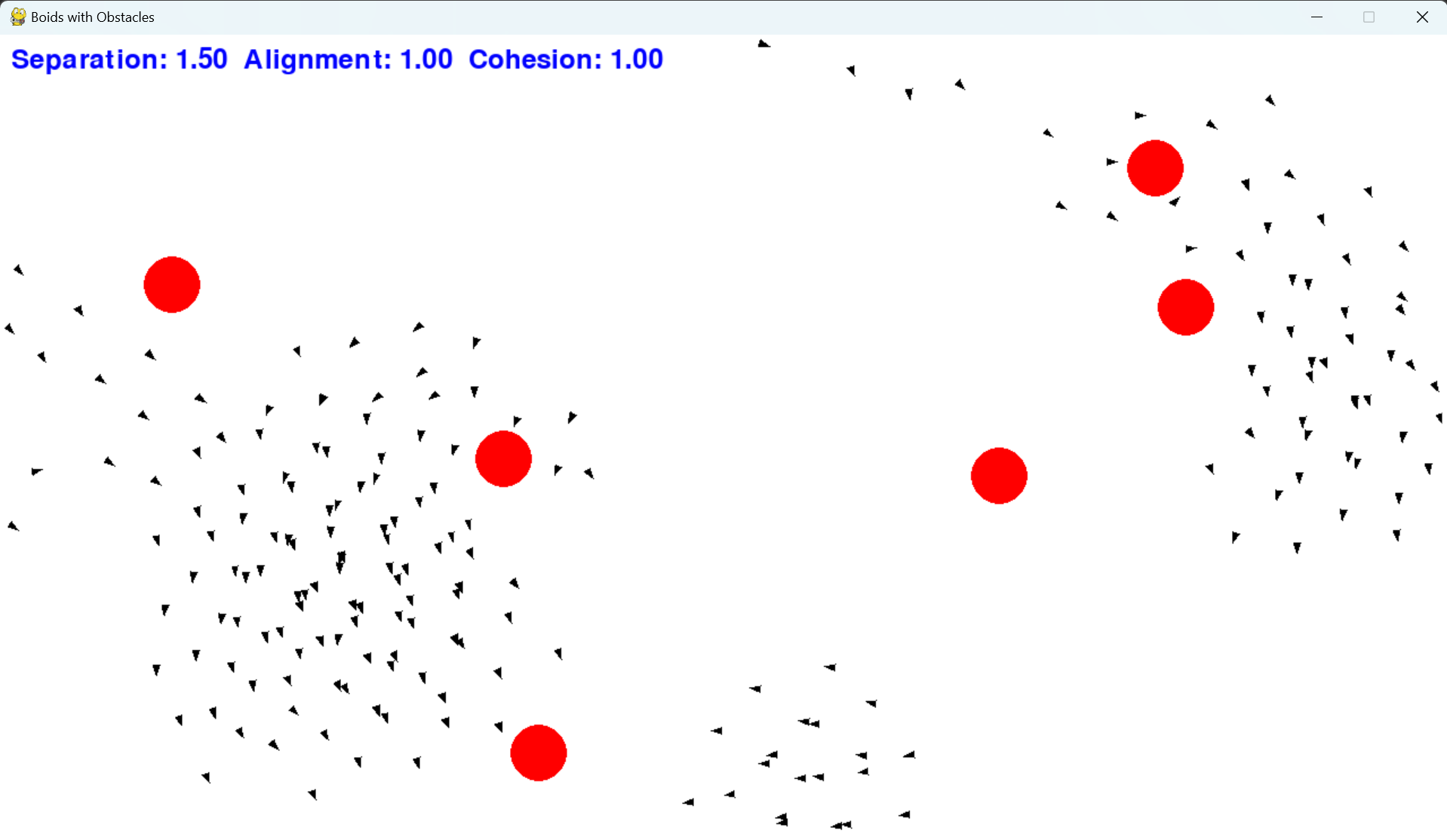}
    \caption{the visualization of the Boid simulation using the code in \ref{sec:data-availability}. The black triangle represents the Boid, and the red circles are obstacles}
    \label{fig:visual}
\end{figure}

The detailed code was posted on GitHub, check \ref{sec:data-availability} for more information. Fig. \ref{fig:visual} illustrates a demo simulation program where we can see that the flocks are formed, and the Boids are moving around. I use \texttt{Pygame} module for visualization.

\subsection{Metrics}
\label{sub:metrics}

To explore the physics behavior of the Boids model, we need to specify some metrics related to motion to capture the natural collective behavior and smooth movement. In this project, I propose 4 metrics as follows:

\begin{itemize}
    \item \textbf{Average speed} of Boids, which assesses the effort Boids exert in their movement compared to the distance they cover in nature
\[ \text{Average Speed (AS)} = \frac{1}{N} \sum_{i=1}^N  \|\mathbf{v}_i \|, \]
    \item \textbf{Flocking radius} is the mean distance of Boids from the center of mass, which indicates how closely the Boids are clustered around their center of mass
\[
\text{Flocking Radius (FR)} = \frac{1}{N} \sum_{i=1}^N \|\mathbf{p}_i - \mathbf{p}_{\text{cm}}\|,
\]
    \item \textbf{Turn rate} measures the average angular change in the velocity direction, which evaluates how often and sharply the Boids change direction
\[
\text{Turn Rate (TR)} = \frac{1}{N} \sum_{i=1}^N |\theta_i|,
\]
    \item \textbf{Dispersion} is the standard deviation of distances of Boids from the center of mass, which quantifies  the dispersion of the Boids, with a focus on identifying outliers
\[
\text{Dispersion (D)} = \sqrt{\frac{1}{N} \sum_{i=1}^N \left(\|\mathbf{p}_i - \mathbf{p}_{\text{cm}}\| - \text{Flocking Radius} \right)^2}
\]
\end{itemize}

, where

\begin{itemize}
    \item $ \mathbf{v}_i $: current velocity of Boid $ i $.
    \item $ \mathbf{p}_i $: current position of Boid $ i $.
    \item $ \mathbf{p}_{\text{cm}} $: Center of mass of the flock, calculated as: $ \mathbf{p}_{\text{cm}} = \frac{1}{N} \sum_{i=1}^N \mathbf{p}_i $.
    \item $ \theta_i $: Angle between the current velocity $ \mathbf{v}_i $ and the previous velocity $ \mathbf{v}_i^{\text{prev}} $ for Boid $ i $.
    \item $ N $: Total number of Boids.
\end{itemize}

\section{Experiments}
\label{sec:exp}

\subsection{Experiemental setup}
\label{sub:exp-setup}

The experiments involve 200 Boids and 6 obstacles, which are randomly initialized for each simulation. To gather statistical results, I conducted three experiments by using the Monte Carlo method, each consisting of 100 simulations with 500 iterations for each \( w_{s} \), \( w_{a} \), or \( w_{c} \). The computer utilized for this experiment is an HP ENVY x360 laptop, equipped with an AMD Ryzen 5 8640HS processor and Radeon 760M graphics, running at 3.50 GHz. The hyperparameters are set up as follows:

\begin{lstlisting}[language=Python, caption=Hyperparameters settings]
# Parameters
SPEED = 3 # The initial speed at which Boids move
MAX_FORCE = 0.5 # The maximum force (F_{max}) that can be applied to a Boid for it to change direction
WANDER_RADIUS = 0.2 # The radius (r_{w}) within which a Boid will wander randomly
AVOID = 10 # The degree (weight w_{o}) to which Boids try to avoid the obstacles
SEPARATION_RADIUS = 50 # The radius within which Boids will begin to experience separation force
ALIGNMENT_RADIUS = 50 # The radius within which Boids will begin to experience alignment force
COHESION_RADIUS = 200 # The radius within which Boids will begin to experience cohesion force
\end{lstlisting}

The small separation and alignment radii, along with the large cohesion radius, represent local and global behaviors, respectively. Boids will look for the moving direction in a far vision while keeping their distance from their neighbors. At an initial state, all Boids have the same speed although their directions are different. 

In this project, the three main parameters are \(w_{s}\), \(w_{a}\), and \(w_{c}\). One of these parameters will be varied for each Monte Carlo experiment, while the other two will remain constant to observe their effect on the model.

\subsection{Experimental Results}
\label{sub:exp-results}
\begin{figure*}[htb!] 
    \centering
        \begin{subfigure}{\linewidth} 
            \includegraphics[width=\linewidth, height = 4.25cm]{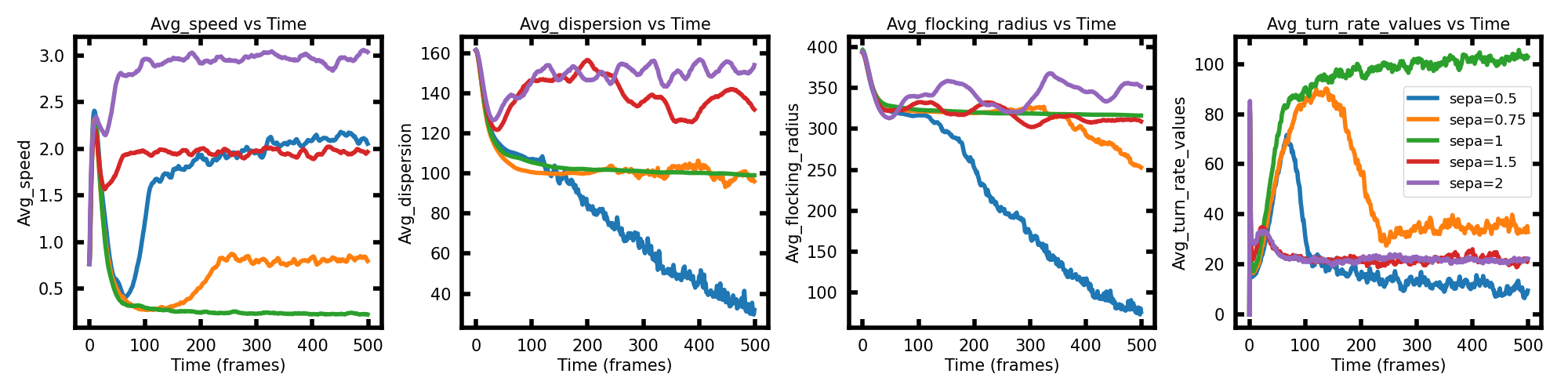}
            \caption{The results for different \( w_{s} \); \(w_{a} = w_{c} = 1\)}
            \label{fig:separation}
        \end{subfigure}
        \begin{subfigure}{\linewidth} 
            \centering
            \includegraphics[width=\linewidth, height = 4.25cm]{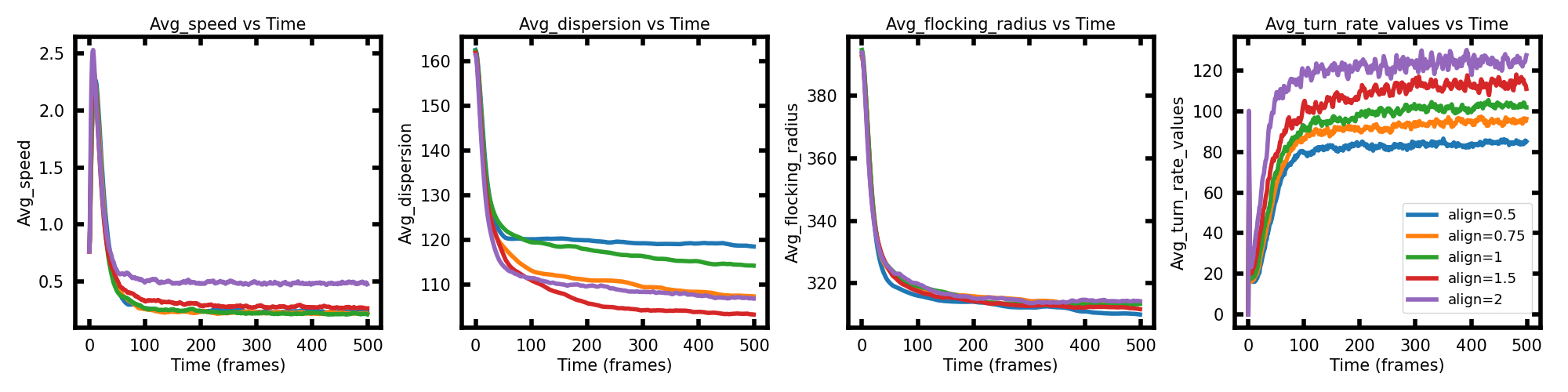}
            \caption{The results for different \( w_{a} \); \(w_{s} = w_{c} = 1\)}
            \label{fig:alignment}
        \end{subfigure}
        \begin{subfigure}{\linewidth} 
            \centering
            \includegraphics[width=\linewidth, height = 4.25cm]{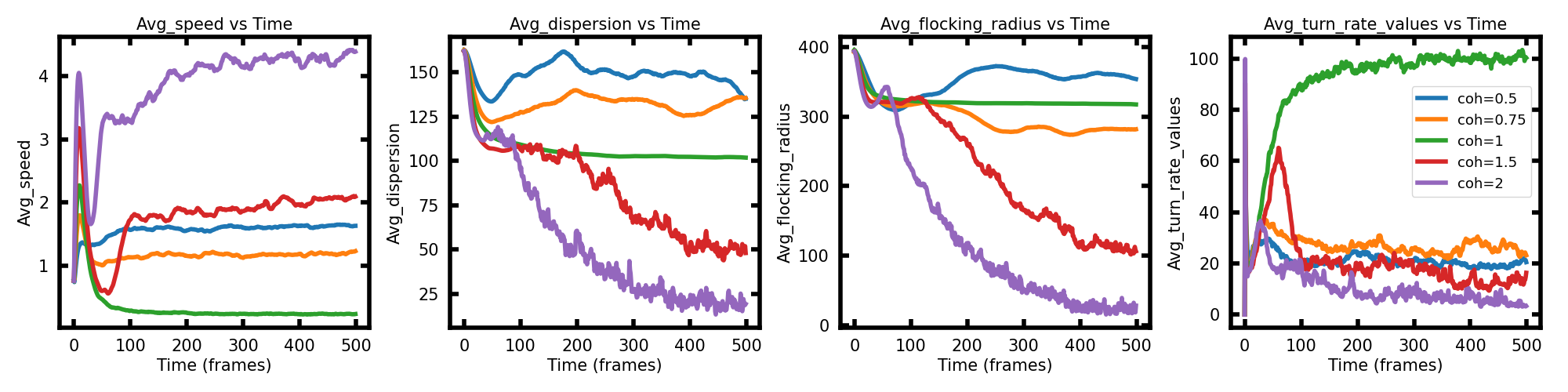}
            \caption{The results for different \( w_{c} \); \(w_{s} = w_{a} = 1\)}
            \label{fig:cohesion}
        \end{subfigure}
    \caption{The results of the experiment with the setting in \ref{sub:exp-setup}. The variation in \(w_{a}\) has minimal impact on the behavior of Boids, as demonstrated in Fig. \ref{fig:alignment}. However, changes in separation (\(w_{s}\)) and cohesion (\(w_{c}\)) significantly influence the size and formation of the flock. Figures \ref{fig:separation} and \ref{fig:cohesion} show that flocks remain small and tightly clustered when \(w_{c}\) exceeds \(w_{s}\), whereas the opposite causes the flock to spread out. Over time, the results further reveal that the velocity of each Boid remains stable, both in magnitude (evidenced by the average speed) and direction (indicated by the turn rate).
}
    \label{fig:result}
\end{figure*}

Fig. \ref{fig:result} depicts the results from using the Monte Carlo method to analyze the collective behavior of the model. While the variation in \(w_{a}\) does not significantly affect the behavior of Boids, as shown in Fig.\ref{fig:alignment}, changes in separation or cohesion do influence the formation of the flock's size. Both Fig. \ref{fig:separation} and Fig. \ref{fig:cohesion} illustrate that flocks are small (as shown in dispersion and flocking radius plots) when \(w_{c} \) exceeds \( w_{s} \) and vice versa. This means the Boids are tightly clustered, rather than spreading out. Besides that, when \( w_{s} = w_{c} \), the Boids are not moving or just fluctuate about their position due to the "wander" rule and the annihilation of separation and cohesion forces. For long-term behavior, the results also indicate that the velocity of each Boid is stable in both magnitude (as shown by the average speed) and direction (as indicated by the turn rate). In case all Boids have the same potential energy, it becomes evident that their stable speed and minimal turning rate allow for smooth movement. This results in the near conservation of kinetic energy during their flight. This principle helps to explain bird migration. Birds often fly in flocks to save energy, enabling them to maintain a stable velocity and cover greater distances. In this model, the Boids' velocity stabilizes rapidly for any combination of \( w_{s} \), \( w_{a} \), and \( w_{c} \). This stabilization occurs approximately after 50 to 100 iterations, as indicated by the saturation of the average speed and average turn rate.

\section{Discussion and Future Works}
\label{sec:discuss}

This project presents a model for non-colliding aggregate motion, similar to the behavior of flocks of birds. The model simulates the actions of each "Boid" independently. Each Boid aims to stay close to its neighbors while also avoiding collisions with other Boids and objects in its environment. While the "migration" properties of the model have been analyzed, there are still some aspects that remain unexplored, such as the unification and splitting of flocks as in \cite{Maruyama2019}. In the future, additional rules can be incorporated into the model to represent more realistic behavior, and new techniques, such as machine learning, can be employed to analyze this behavior.

\section{Contact me}
\label{sec:contact}

You can contact me by email for more information: \href{mailto:quocchuo@buffalo.edu}{quocchuo@buffalo.edu}.
    
\section{Data availability}
\label{sec:data-availability}

All codes and results of this project are posted on:
\begin{center}
\url{https://github.com/ChuongQuoc1413017/Multi_Agent_Simulation/blob/main/Boids_Simulation-Final%20Version.ipynb}
\end{center}

\bibliographystyle{unsrt}  
\bibliography{references}

\end{document}